\documentclass[journal]{IEEEtran}

\usepackage{cite}
\usepackage{amsmath}
\usepackage{array}
\usepackage{booktabs}
\usepackage{graphicx}
\usepackage{float}
\usepackage{subfig}
\usepackage{color}
\usepackage{amsfonts}

\begin{document}

\title{An Edge Computing Paradigm for Massive IoT Connectivity over High-Altitude Platform Networks}

\author{Malong~Ke,
Zhen~Gao,~\IEEEmembership{Member,~IEEE,}
Yang~Huang,~\IEEEmembership{Member,~IEEE,}
Guoru~Ding,~\IEEEmembership{Senior Member,~IEEE,}
Derrick~Wing~Kwan~Ng,~\IEEEmembership{Fellow,~IEEE,}
Qihui~Wu,~\IEEEmembership{Senior Member,~IEEE,}
and Jun~Zhang,~\IEEEmembership{Member,~IEEE}}

\maketitle

\begin{abstract}
With the advent of the Internet-of-Things (IoT) era, the ever-increasing number of devices and emerging applications
have triggered the need for ubiquitous connectivity and more efficient computing paradigms.
These stringent demands have posed significant challenges to the current wireless networks and their computing architectures.
In this article, we propose a high-altitude platform (HAP) network-enabled edge computing paradigm to tackle the key issues
of massive IoT connectivity.
Specifically, we first provide a comprehensive overview of the recent advances in non-terrestrial network-based
edge computing architectures.
Then, the limitations of the existing solutions are further summarized from the perspectives of the network architecture,
random access procedure, and multiple access techniques.
To overcome the limitations, we propose a HAP-enabled aerial cell-free massive multiple-input multiple-output network to
realize the edge computing paradigm, where multiple HAPs cooperate via the edge servers to serve IoT devices.
For the case of a massive number of devices, we further adopt a grant-free massive access scheme to guarantee low-latency
and high-efficiency massive IoT connectivity to the network.
Besides, a case study is provided to demonstrate the effectiveness of the proposed solution.
Finally, to shed light on the future research directions of HAP network-enabled edge computing paradigms, the key challenges
and open issues are discussed.
\end{abstract}

\vspace{-1mm}
\section{Introduction}
\label{Sec.I}

The Internet-of-Things (IoT) is playing a major role across a variety of vertical sectors by generating tremendous cost savings
and new revenue streams.
As a result, it is the common consensus that the next-generation wireless networks should efficiently and reliably support
massive IoT connectivity with guaranteed quality of service (QoS)~\cite{Porambage_Survey'18}.
A key feature of IoT is the low-cost devices with limited computational and storage capabilities,
thus it is challenging for these devices to perform computationally heavy and latency-sensitive tasks locally.
Traditionally, the IoT applications are executed in a centralized manner, i.e., via cloud computing,
where the tasks of IoT devices are collected at a remote centralized cloud server for further processing~\cite{Mach_Survey'17}.
Nevertheless, this conventional computing approach introduces a significant transmission delay for the devices'
computation offloading.
As a promising alternative, the recently proposed edge computing paradigm aims at extending cloud computing capabilities
to the edge of radio access networks, hence alleviating the backhaul congestion, reaping a faster response,
and offering a better QoS for the devices~\cite{Porambage_Survey'18},~\cite{Mach_Survey'17}.

In traditional cellular network-based edge computing architectures, the inherent defects of terrestrial
infrastructures would be the severely limiting factors to achieve the huge potential gains brought by IoT~\cite{Qiu_WC'19}.
Specifically, the unprecedented escalating data traffic and processing demands will easily overload terrestrial
base stations (BSs), thus jeopardizing the system performance.
Moreover, it is expensive to deploy terrestrial BSs in remote areas with low population density, difficult terrain,
and the lack of infrastructures like a power grid.
In addition, some emergency scenarios such as natural disasters can also interrupt terrestrial connectivity.

As supplements to terrestrial networks, non-terrestrial networks (NTNs) can offer effective coverage to
the areas where terrestrial BSs are overloaded or unavailable~\cite{{Qiu_WC'19}}.
NTNs usually include satellite constellations at different Earth's orbits, high-altitude platform (HAP) networks,
and unmanned aerial vehicle (UAV) networks.
In contrast with the frequently changed position, exceedingly large delay, and expensive operational costs of satellites,
HAPs (e.g., airship and balloon) positioned in the stratosphere (about 20 km altitude) have the advantages of providing
stable, low-latency, and affordable services for IoT.
Moreover, compared with UAVs, HAPs exploiting both buoyancy and solar energy have longer flight endurance,
larger maximum payload capacity, and wider service coverage.
Therefore, integrating edge servers into the HAP network is a promising solution to support communication, computing,
and storage operations for IoT devices working around the clock, which are distributed across the ground, ocean, and air.
In HAP-based edge computing networks, an efficient massive access scheme is vital for IoT devices to request communication
and computation services from the network.
In fact, various industrial leading companies, e.g., Samsung, has predicted that the number of connected devices will reach
500 billion by 2030, which is about 59 times larger than the expected world population by that time \cite{Porambage_Survey'18}.
In this context, traditional grant-based random access protocols will suffer from an extremely high access latency
\cite{Ke_JSAC'20}, and a more efficient massive access scheme is indispensable.

The rest of this article is organized as follows.
In the following sections, we first present the state-of-the-art studies on NTN-based edge computing systems and summarize the
limitations of existing solutions.
Then we propose a HAP-enabled aerial cell-free massive MIMO network to realize the edge computing paradigm,
where a grant-free massive access scheme is further developed to accommodate a massive number of devices.
Finally, the open research issues are discussed and the article is concluded.

\vspace{-1mm}
\section{State of the Art}
\label{Sec.II}

\begin{figure*}[!t]
\centering
\includegraphics[width=16cm]{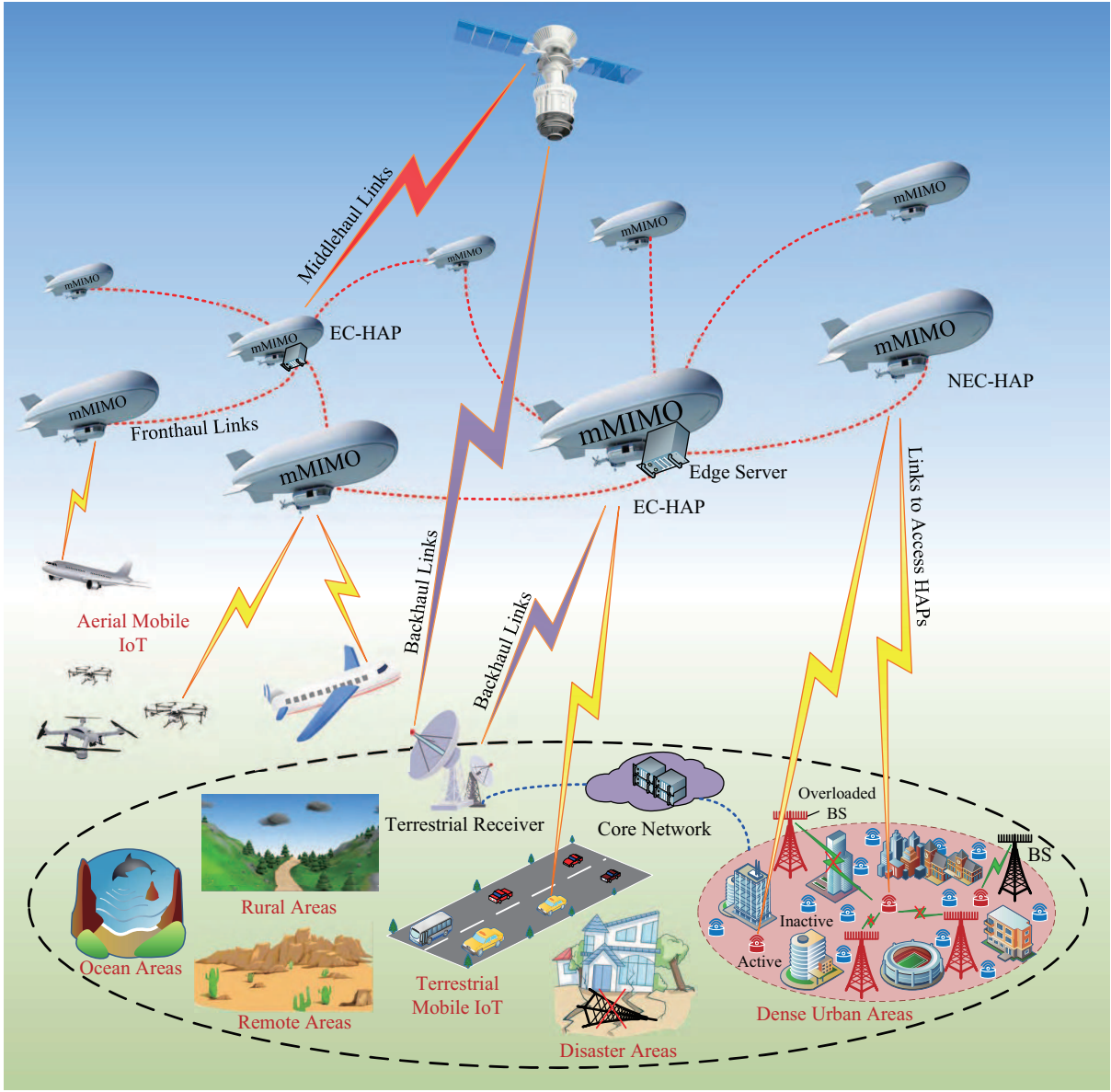}
\caption{\small{The proposed edge computing paradigm for massive IoT connectivity, where the HAP-enabled aerial cell-free
massive multiple-input multiple-output (mMIMO) network is employed and the potential application scenarios are illustrated.}}
\label{Fig1}
\vspace*{-3mm}
\end{figure*}

In this section, we first present the potential application scenarios of HAP-enabled edge computing networks,
as shown in Fig. \ref{Fig1}.
Then, we further provide an overview of the ongoing research efforts in the fields of NTN-based edge computing architectures
and HAP-based wireless networks.

\subsection{Application Scenarios}
\label{Sec.II-A}

\begin{itemize}
\item{\textbf{Improve the QoS for hotspot areas:}
For some grand gatherings, such as concerts and Olympic games, there will be a dramatic increase of devices temporarily,
which results in a massive volume of data and computing tasks.
On the other hand, with the development of IoT, the emerging applications pose more strict signal processing requirements
on the BS.
These may overload the terrestrial edge computing networks and increase the outage probability, thus degrade the users'
quality of experience.
In this case, HAP-enabled edge computing networks can be employed to assist the terrestrial systems for executing
computing tasks.
Particularly, the overloaded terrestrial BSs can offload part of their computing tasks to the edge servers carried by HAPs,
such that the devices failing to access the terrestrial BSs can directly establish the connectivity with HAP networks.}

\item{\textbf{Provide ubiquitous connectivity for remote areas:}
The remote areas usually have low population density, low expected revenue, and a lack of terrestrial infrastructures.
These facts reduce the incentive of commercial companies to invest and operate connectivity networks in these areas.
Consequently, roughly half the world population remains unconnected or poorly connected, causing a sharp digital divide.
Meanwhile, the constant monitoring of the areas such as wildernesses, deserts, and complex terrains is essential for
smart agriculture and environment protection, where a low-cost solution providing a wide coverage is indispensable.
For these areas, direct network access via satellite systems is a potential solution.
Yet, the high operational cost and relatively low data rates are the major limiting factors to realize ubiquitous
connectivity.
As a remedy, the HAP-enabled edge computing network can serve as an affordable solution which is capable of delivering
a competitive QoS in these areas to those in urban areas.}

\item{\textbf{Emergency services for disaster areas:}
During natural disasters, terrestrial edge computing networks may be damaged.
However, many urgent rescue or reconstruction related tasks need to be executed in time.
In particular, the high maneuverability of HAPs allows them to travel a vast distance and switch between
moving and strategically hovering over a specific area.
Thus they can be rapidly deployed for establishing a reliable temporary network for communication and computing
in disaster zones.}

\item{\textbf{Support mobile IoT applications:}
For mobile IoT applications, such as autonomous driving and UAV communications, the devices usually move at a high speed.
Due to the smaller cellular coverage, mobile IoT applications in terrestrial networks  require a frequent BS handover.
By leveraging the ultra-wide coverage of HAPs, the required number of handovers can be significantly reduced,
which improves the reliability of communication.}
\end{itemize}
For the aforementioned application scenarios, the HAP network, i.e., aerial radio access network,
usually directly connects to the terrestrial core network via backhaul links for communication services.
This will significantly reduce the propagation delay and operational cost.
However, in some extreme cases where the HAP network fails to establish direct communication links with terrestrial
infrastructures, the satellites with global coverage have to be employed to fill in the gap between the HAP network and
the core network, and the additional middlehaul links are required to connect the HAP network and the satellites.

\subsection{Recent Advances}
\label{Sec.II-B}

Both research and industry communities have made great efforts to accelerate the advancement of edge computing-based IoT
ecosystems, with the focus on integrating edge computing into terrestrial cellular networks.
For instance, the researchers have proposed various solutions to address the essential issues in edge computing, such as
computation offloading, resource allocation, and overall delay minimization problems
~\cite{{Porambage_Survey'18}, {Mach_Survey'17}}.
Apart from the academic researches, European Telecommunications Standards Institute has been deeply involved in
the standardization of edge computing by launching an Industry Specification Group since December 2014~\cite{Mach_Survey'17}.

Due to the evolution in communication and aerospace technologies, NTN-based edge computing architecture
has been identified as a viable solution to provide ubiquitous coverage and to execute computation-intensive applications
for massive IoT devices~\cite{{Zhou_WC'20}, {Cheng_CM'18}}.
In particular, most recent studies employed UAV as the aerial platform to carry edge servers, which can either assist
the terrestrial BSs or directly connect the devices to provide various services.
For example, in~\cite{{Cheng_CM'18},{Cheng_JSAC'19}}, the authors established an air-ground integrated edge computing network,
where UAVs are flexibly deployed and scheduled to assist the communication, caching, and computing of terrestrial networks.
Considering the UAVs can provide direct communication links for devices, the minimization of the mobile energy consumption and
the computation rate maximization problems were further investigated~\cite{{Jeong_TVT'18}, {Zhou_JSAC'18}}.

In addition to UAVs, satellites and HAPs are the other two promising types of aerial platforms for realizing
NTN-based edge computing architectures.
Compared with UAVs and satellites, HAPs have a much better tradeoff between their merits and faults, involving service endurance,
coverage area, air-ground channel stability, overall transmission and processing delay, as well as the launch and recovery costs.
The advantages of HAPs over UAVs and satellites are detailed as follows:
\begin{itemize}
\item{In contrast to UAVs, HAPs (also known as stratospheric satellites) operating at a higher altitude typically provide
coverage with larger geographic areas.
The radius of HAPs' coverage can be up to 200 km, while that for UAVs is just within several kilometers.
Particularly, a handful of HAPs could even cover a whole country, for instance, Japan can be covered by 16 HAPs with
an elevation angle of $10^{\circ}$ while Greece can be covered by 8 HAPs~\cite{Cao_JSAC'18}.
Moreover, unlike UAVs with limited onboard energy, HAPs have a larger payload capability (up to 500 kg) and much longer
endurance (up to several years), which make them the more appealing candidate to carry edge servers and to provide persistent
computing services.}

\item{Owing to their inherent characteristics, low Earth orbit (LEO) satellites and medium Earth orbit (MEO) satellites
always remain in a high-speed movement.
This leads to the fast time-varying channel propagation environments between LEO/MEO satellites and terrestrial devices.
On the contrary, with the benefits of the stratosphere, HAPs can stay at a stationary point for a long time, and toggle
between moving and strategically hovering at a will.
Hence the stable HAP-ground channel can be achieved.}

\item{The flight altitude of HAPs is less than $0.1\%$ of the altitude of geostationary Earth orbit (GEO) satellites
(about 36,000 km), and less than $6\%$ of the altitude of LEO satellites (500 - 2,000 km).
This attainable distance allows HAPs to use light gas (such as hydrogen and helium) or solar power to efficiently and
sustainably power their flight, which significantly reduces the launching and recovery costs for flexible deployment
or regular maintenances.
Meanwhile, due to the shorter distance, the lower transmit power and propagation delay can be achieved in HAP networks.}
\end{itemize}
Despite the superiorities of HAPs, the literature along the direction of HAP network-enabled edge computing paradigm
for massive IoT connectivity is still limited.
Recently, the investment in the HAP industry by Google, through its LOON project, has brought back the attention to HAPs,
motivating both the industry and academia to invest in and study HAP-based wireless networks.
Moreover, the other HAP projects, including the Zephyr of Airbus, the Aquila of Facebook, the StratoBus of Thales,
and the Flying Cell on Wings of AT$\&$T company, also pave the way towards HAP-enabled wireless communication and
edge computing networks \cite{Qiu_WC'19}.

\begin{figure*}[!t]
\centering
\subfloat[]{\includegraphics[width=2.2in]{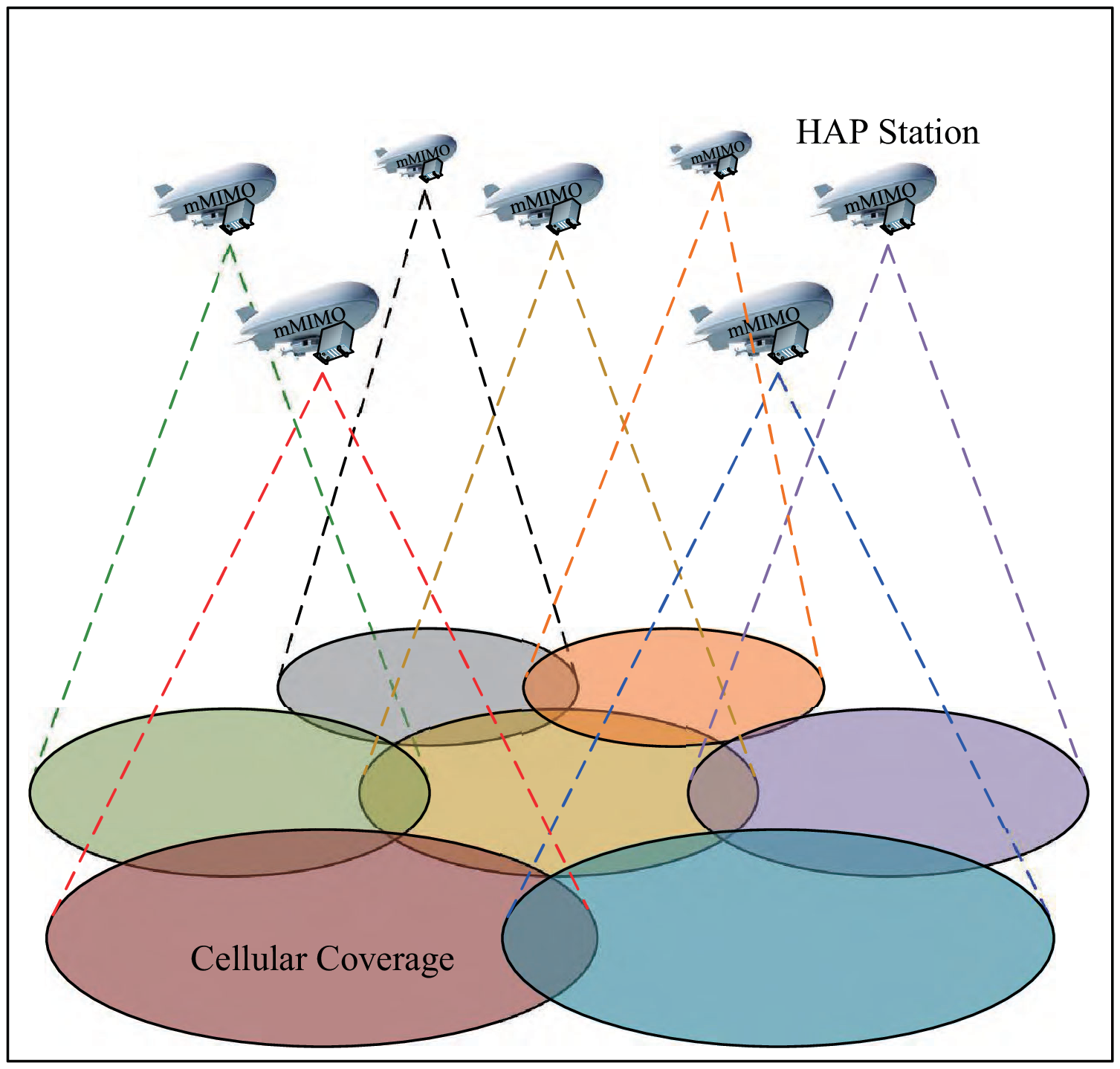}}
\hfil
\subfloat[]{\includegraphics[width=2.2in]{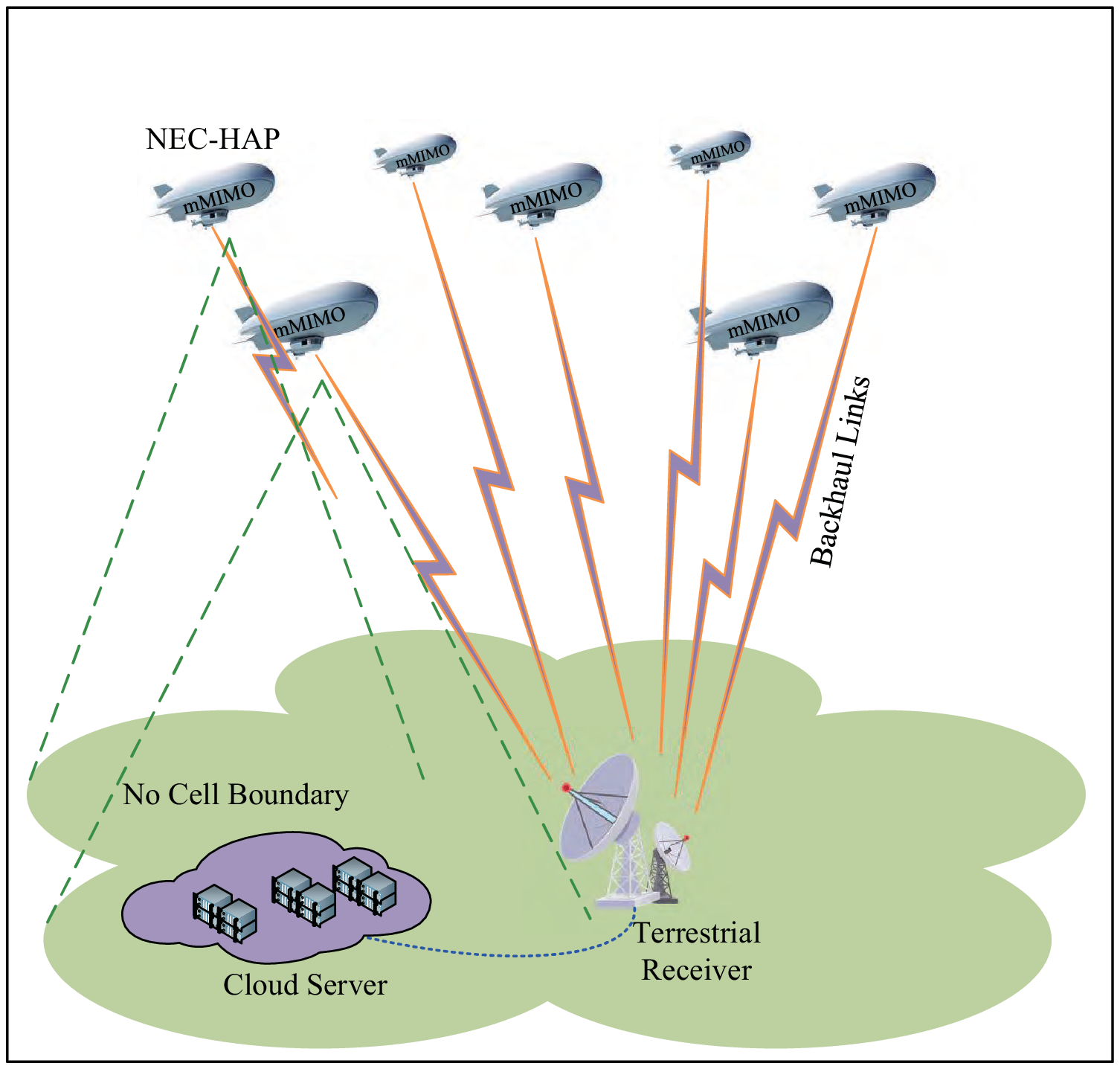}}
\hfil
\subfloat[]{\includegraphics[width=2.2in]{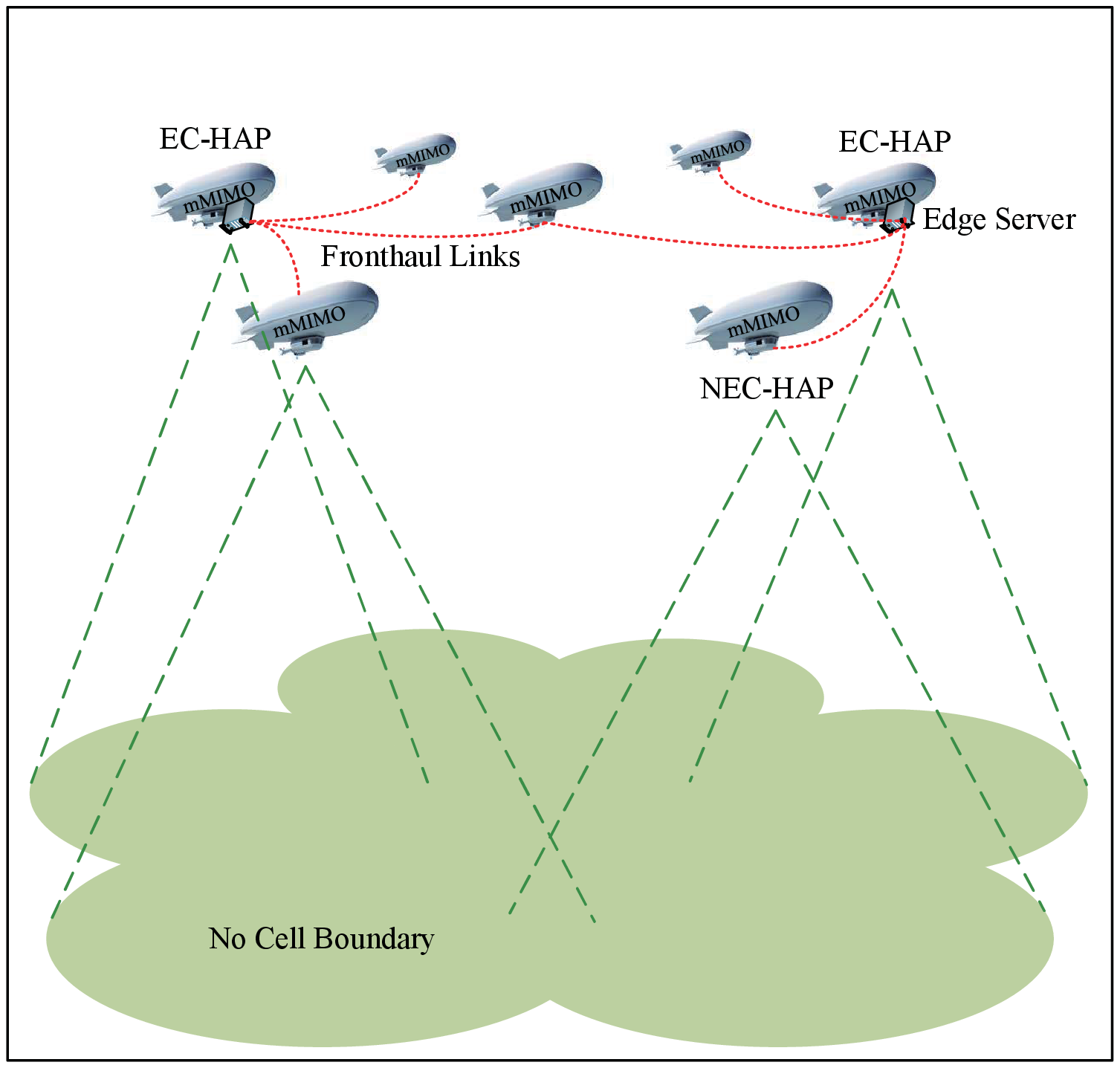}}
\caption{\small{(a) HAP-based cellular network; (b) HAP-based cell-free massive MIMO network for {\em cloud computing};
                (c) HAP-based cell-free massive MIMO network for {\em edge computing}.}}
\label{Fig2}
\vspace*{-3mm}
\end{figure*}

\section{Limitations of Existing NTN-Based Edge Computing Solutions}
\label{Sec.III}

The research of the NTN-based edge computing paradigm is still in its infancy.
In fact, the massive number of devices and their vast distribution have posed significant challenges to the existing
schemes in realizing massive IoT connectivity.
In this section, in terms of the network architecture, random access procedure, and multiple access techniques,
we discuss the major limitations of existing NTN-based edge computing solutions for supporting massive IoT connectivity.

\subsection{Network Architecture}
\label{Sec.III-A}

The network architecture is the foundation of NTN-based edge computing systems.
Previous work employed various aerial platforms as the aerial BSs to provide cellular coverage for terrestrial devices,
where only a single-cell scenario is investigated~\cite{Zhou_WC'20}, \cite{Jeong_TVT'18}, or multiple aerial BSs
do not cooperate with each other~\cite{Cheng_CM'18},~\cite{Cheng_JSAC'19}, as illustrated in Fig.~\ref{Fig2}(a).
Due to the heterogeneous path loss~\cite{Ke_JSAC'20}, a main drawback of the traditional cellular networks is the
inconsistent user experience between the cell center and the cell edge.
This phenomenon will become more severe in NTN, as the cell radius of aerial BSs is usually much larger than that of
terrestrial BSs.
Moreover, since the power-limited devices are widely distributed, a large number of aerial BSs should be deployed to
guarantee the availability of ubiquitous connectivity and edge computing services.
In this context, multiple access interference may arise between multiple uncoordinated aerial BSs utilizing the same
radio spectrum.
To avoid inter-cell interference and offer uniformly good QoS, an efficient cooperation strategy among
multiple aerial BSs is indispensable.

\subsection{Random Access Procedure}
\label{Sec.III-B}

The connection procedure of random access affects the network performance in terms of latency, energy consumption,
and the number of supported devices.
Most previous work considered human-type communication scenarios with a relatively small number of devices,
where traditional grant-based random access protocols can be directly applied~\cite{Zhou_WC'20},~\cite{Jeong_TVT'18}.
For instance, the physical random access channel (PRACH) protocol of 4G Long Term Evolution is a typical representative
of grant-based random access protocols, which requires a 4-step interaction between devices and the BS for
access scheduling and communication resources granting.
When applying PRACH to massive IoT connectivity, multiple devices may choose the same preamble to access the BS as
the number of devices is large while the number of assigned preambles is limited.
This will lead to a severe access collision, and the complicated contention resolution for eliminating
the access interference among colliding devices will introduce a high access latency.
Therefore, the simplification of access scheduling and the circumvention of complicated contention resolution
are essential for providing low-latency communication and computation services for massive IoT devices.

\subsection{Multiple Access Techniques}
\label{Sec.III-C}

The multiple access techniques used for payload data transmission of multiple devices can be classified into two categories:
orthogonal multiple access (OMA) and non-orthogonal multiple access (NOMA).
In OMA, the radio resources are divided into multiple orthogonal resource blocks (RBs) to convey the data of different
devices, which can avoid inter-device interferences and facilitate the BS to achieve reliable data detection with
a simple receiver design.
In UAV-enabled edge computing networks, the typical OMA schemes such as time/frequency division multiple access
have been extensively studied~\cite{{Cao_JSAC'18}, {Lyu_TWC'18}}.
However, the number of the accommodated devices is limited by the number of orthogonal RBs.

Different from OMA, NOMA allows a larger number of simultaneously served devices than the number of occupied RBs,
as the signals of different devices can be transmitted via the non-orthogonal RBs.
The core idea of NOMA lies in distinguishing different devices from the overlapping radio resources based on the devices'
diversity in a specific domain, such as different power levels in the power domain NOMA and different codewords in
the code domain NOMA.
The authors in~\cite{Liu_JSAC'20} validated that NOMA-based edge computing outperforms OMA-based edge computing in terms of
the computation performance gain.
However, for traditional NOMA schemes (such as power-domain and code-domain NOMA), the design of devices' diversity
(i.e., power levels and  codebook, respectively) and the computationally efficient algorithm for multiple devices'
data detection can be extremely complicated when considering a massive number of devices.

\section{Proposed Aerial Edge Computing Paradigm for Massive IoT Connectivity}
\label{Sec.IV}

To overcome the aforementioned limitations, this section develops an advanced edge computing paradigm
tailored for massive IoT connectivity, which can provide the devices with ubiquitous, affordable, and low-latency
communication and computation services.

\subsection{HAP-Enabled Aerial Cell-Free Massive MIMO Network}
\label{Sec.IV-A}

Considering the attractive benefits of HAPs, the proposed edge computing network architecture is illustrated in
Fig. \ref{Fig1} and Fig. \ref{Fig2}(c), where a large number of HAPs cooperate in the stratosphere to provide
uniformly good services anywhere and anytime.
The HAPs, equipped with a massive number of antennas, can be classified into two categories in terms of their
functionalities: HAPs with and without edge computing servers, denoted as EC-HAP and NEC-HAP,
respectively.
NEC-HAP is designed as an aerial access point for receiving and transmitting signals, where only simple
communication antennas and related radio frequency chains are required.
While the EC-HAP carries an additional edge server onboard, hence can work as a processing unit for executing various
computing tasks, including complicated signal detection, large-volume data processing, information fusion, etc.
Moreover, NEC-HAPs are connected to several adjacent EC-HAPs via high-speed line-of-sight (LoS) fronthaul links for
distributed HAP cooperation.
We can further explain this distributed cooperation strategy from an EC-HAP-centric perspective.
For a specific EC-HAP, its edge server will collect the signals received locally and from $(N_{co} - 1)$ nearest HAPs
for joint processing, where $N_{co}$ is the number of HAPs for cooperation.
By performing coherent signal processing across geographically distributed HAP antennas, this network architecture,
which is an aerial cell-free massive MIMO network, can provide uniformly good service to all devices in the coverage.
In this way, the terrestrial devices can directly access these HAPs to request communication services or offload their
computation tasks to the edge server.

In HAP-based {\em cellular} networks, each HAP station operates independently to serve its own cell's devices,
as shown in Fig. \ref{Fig2}(a).
In this case, the inter-cell interference is a severely limiting factor for high-quality services, and the QoS
at the cell edge is usually poor.
While in the proposed {\em cell-free} network architecture, due to the effective HAP cooperation, the concept of
cell boundary does not exist, thus the inter-cell interference can be avoided and the uniformly good QoS can be achieved.
Moreover, since only a small number of HAPs are equipped with additional edge servers, the proposed architecture can
significantly reduce the HAPs' cost for their large-scale deployment.

On the other hand, compared with conventional cell-free massive MIMO networks adopting a centralized {\em cloud computing}
paradigm, as shown in Fig. \ref{Fig2}(b), the proposed network architecture considers a distributed {\em edge computing}
paradigm, which is capable of reducing the burden on backhaul links and having a much lower overall latency.
First, in cloud computing paradigm, the signals received at all HAPs are transferred to a cloud server for joint processing,
which poses great challenges on the backhaul links between HAPs and the cloud server, especially for the wireless backhaul
with limited capacity.
Second, the edge servers are deployed at the edge of the network, while the cloud server in cloud computing
is usually very far away from the devices.
This results in a much smaller propagation delay for edge computing than that for cloud computing.
Moreover, cloud computing requires the information to pass through several networks including the radio access network,
the backhaul network, and the core network, where the traffic control and routing can contribute to excessive delays.
Besides, with the rapid development of processors, edge servers are powerful enough for executing highly sophisticated
computing programs with a competitive computation latency compared to cloud computing.
Therefore, edge computing reaps almost the benefits of cloud computing but only leads to a lower overall latency.

\subsection{Grant-Free Massive Access Scheme}
\label{Sec.IV-B}

Adopting the grant-free massive access scheme proposed in~\cite{{Ke_TSP'20}, {Liu_TSP'18}},
the random access procedure in the proposed network can be summarized as follows:
\emph{Step 1)} Without any access scheduling, all active devices directly transmit their non-orthogonal pilot
sequences and the subsequent payload data in the uplink;
\emph{Step 2)} NEC-HAPs collect the signals received from all active devices and deliver them to the adjacent
EC-HAPs;
\emph{Step~3)} By jointly processing the signals received at the EC-HAPs locally and their cooperative NEC-HAPs,
the edge servers perform active device detection (ADD) and channel estimation (CE) for the whole network.
Without complex access scheduling and contention resolution, the resulting low access latency makes it an appealing
solution to support massive IoT connectivity.
However, since the signals of all active devices are overlapped on the same radio resources, the ADD and CE
would be the new challenges.

\subsubsection{Frame Structure}
\label{Sec.IV-B-1}

\begin{figure*}[!t]
\centering
\includegraphics[width=6in]{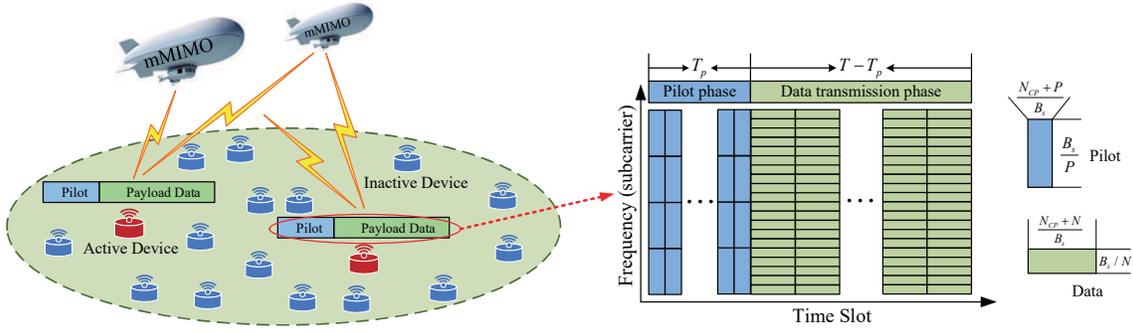}
\caption{\small{The proposed grant-free massive access scheme, where the DFT lengths of pilot phase and
data transmission phase are different.}}
\label{Fig3}
\vspace*{-3mm}
\end{figure*}

\begin{figure*}[!t]
\centering
\subfloat[]{\includegraphics[width=3in]{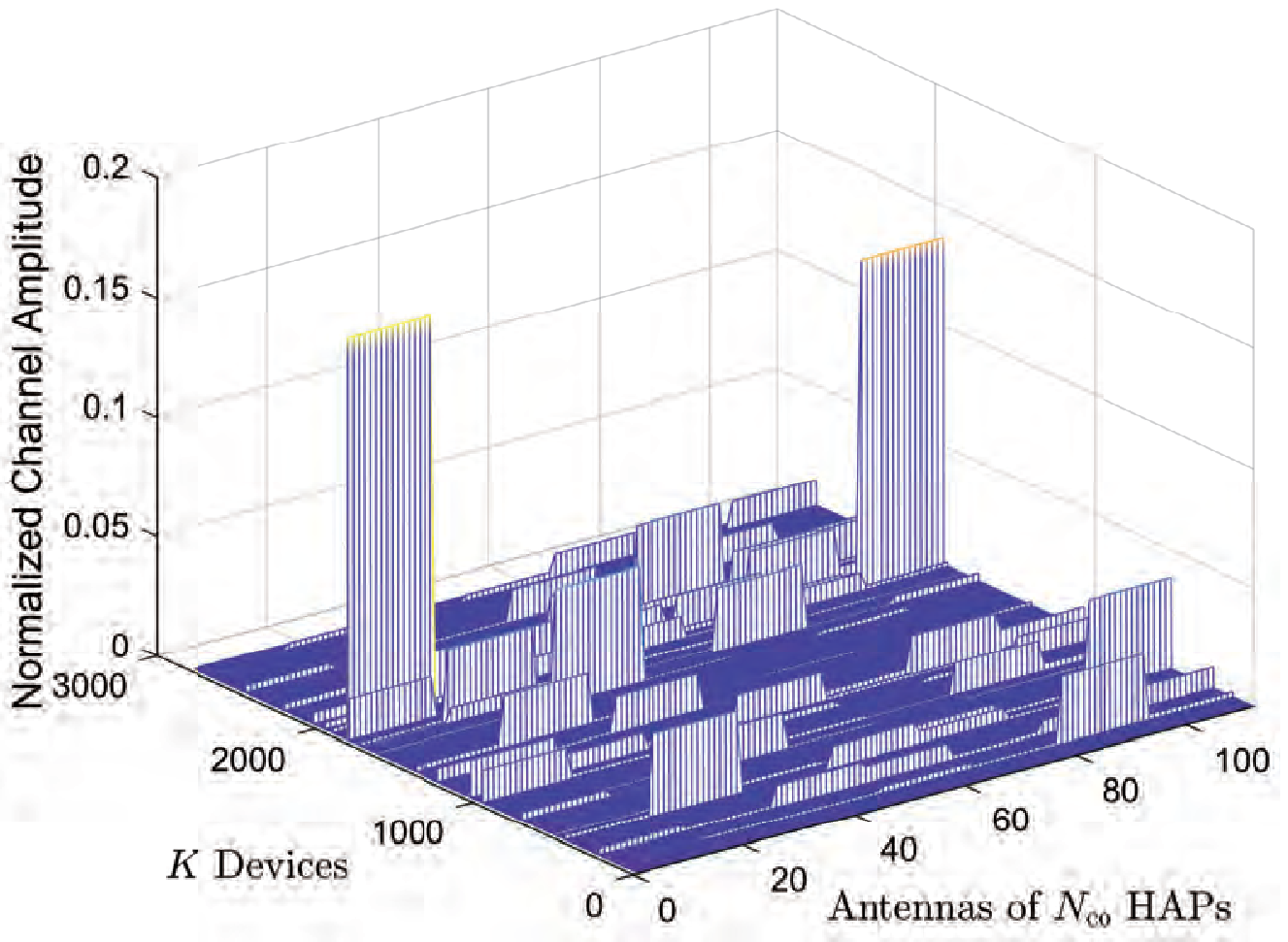}}
\hfil
\subfloat[]{\includegraphics[width=3in]{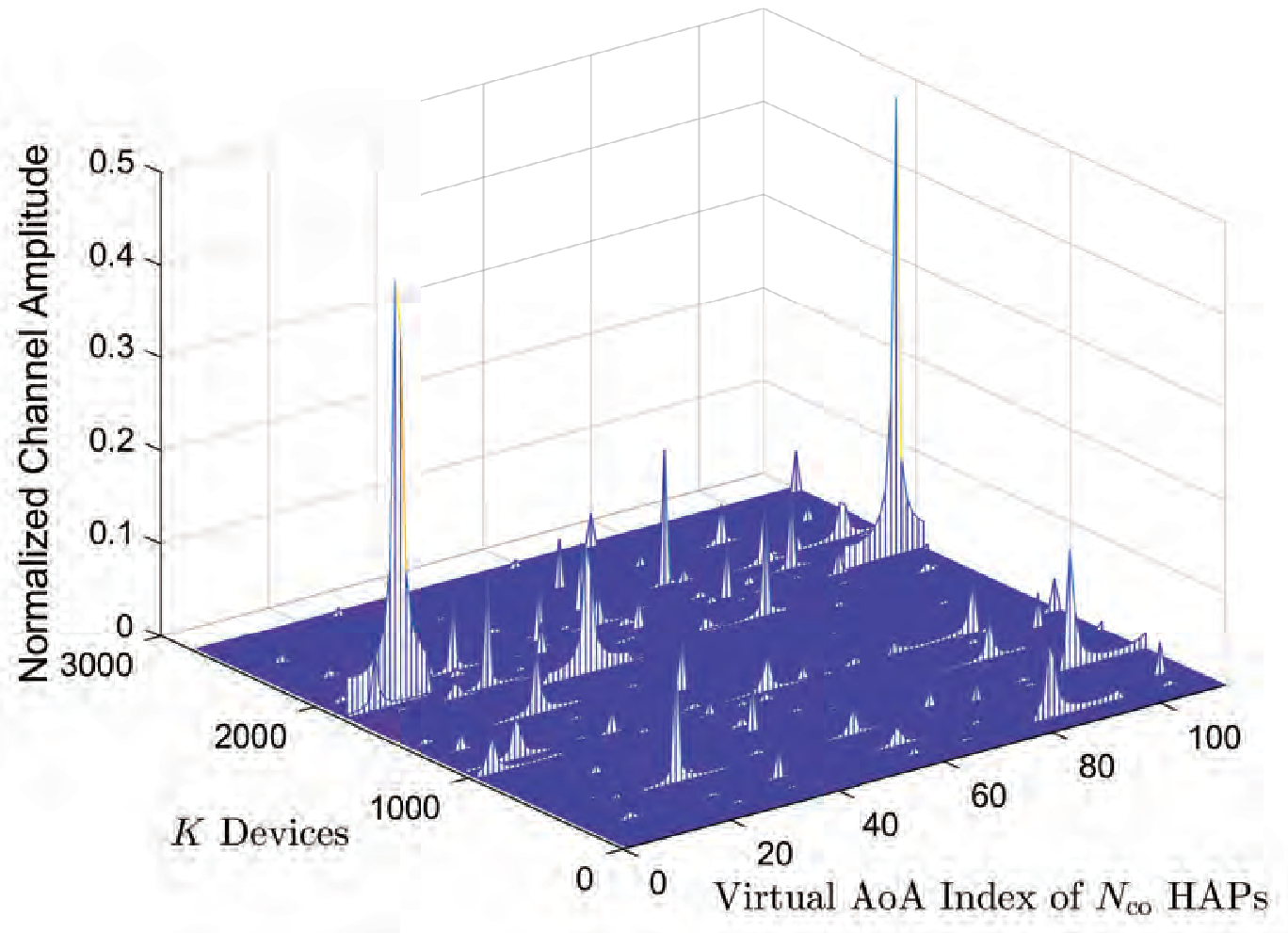}}
\caption{\small{(a) Approximate common sparsity of the spatial-domain channel matrix;
(b) Enhanced sparsity of the angular-domain channel matrix.}}
\label{Fig4}
\vspace*{-3mm}
\end{figure*}

In \emph{Step 1}, the frame structure illustrated in Fig. \ref{Fig3} is employed to reduce
the pilot transmission time.
Here, cyclic prefix (CP)-orthogonal frequency division modulation (OFDM) is adopted and
the length of CP is denoted by ${N_{CP}}$.
A frame is divided into the pilot phase and the data transmission phase in the time domain.
For the pilot phase, we consider the discrete Fourier transform (DFT) length of an OFDM symbol is $P = N_{CP}$,
thus each pilot symbol's duration is $(N_{CP} + P)/B_s$, where $B_s$ is the two-sided bandwidth.
While for the data transmission phase, we consider a different DFT length of $N \gg P$ for a high data rate.
By contrast, the conventional frame structure adopts the same DFT length $N$ in both pilot and data phases,
which leads to a pilot symbol duration of $(N_{CP} + N)/B_s$~\cite{Ke_TSP'20}.
Therefore, the conventional frame structure requires a longer pilot transmission time than the proposed one.

\subsubsection{Successive Interference Cancellation (SIC)-Based ADD and CE Algorithm}
\label{Sec.IV-B-2}

In \emph{Step 3}, for the $p$th pilot subcarrier, the signals received at the $i$th EC-HAP can be expressed as
\begin{equation}\label{Eq1}
\begin{aligned}
{\bf Y}_{p,i} &= {\bf S}_p\left[{\bf H}_{p,{\cal B}_i\left(1\right)}, {\bf H}_{p,{\cal B}_i\left(2\right)}, \cdots,
                {\bf H}_{p,{\cal B}_i\left(N_{co}\right)}\right] + {\bf N}_{p,i}\\
              &= {\bf S}_p{\tilde {\bf H}}_{p,i} + {\bf N}_{p,i},
\end{aligned}
\end{equation}
where ${\bf S}_p = \left[{\bf s}_{p,1}, {\bf s}_{p,2}, \dots, {\bf s}_{p,K}\right] \in {\mathbb C}^{T_p \times K}$,
${\bf s}_{p,k}$ is the pilot sequence of the $k$th device, ${\cal B}_i$ denotes the set of HAPs
cooperating at the $i$th EC-HAP, ${\cal B}_i(n)$ is the $n$th element of ${\cal B}_i$,
${\bf H}_{p,{\cal B}_i\left(n\right)} \in {\mathbb C}^{K \times N_r}$ is the MIMO channel matrix between all devices
and the ${\cal B}_i(n)$th HAP, and ${\bf N}_{p,i}$ is the noise.
Due to the sporadic traffic of devices and the heterogeneous path loss, the spatial-domain access channel matrix
${\tilde {\bf H}}_{p,i}$ exhibits the approximate common sparsity, as illustrated in Fig. \ref{Fig4}(a),
which can be exploited to improve the ADD accuracy~\cite{Ke_TSP'20}.
Meanwhile, by representing the massive MIMO channels in the angular domain, the angular-domain access channel matrix
has the enhanced sparsity, cf. Fig. \ref{Fig4}(b).
This is due to the high likelihood of the presence of LoS links between terrestrial devices and stratosphere HAPs.
In fact, the enhanced sparsity can be used for refining the estimated channel state information (CSI)~\cite{Ke_TSP'20}.

To effectively leverage the aforementioned sparsity properties for improved massive access performance,
we employ the SIC-based ADD and CE algorithm, which consists of three modules~\cite{Ke_JSAC'20}.
Module A detects active devices by estimating ${\tilde {\bf H}}_{p,i}, \forall p,i$, based on
the spatial-domain channel model (1).
Module B refines the CSI estimates of identified active devices based on the angular-domain channel model.
Module C executes the SIC procedure via cancelling the signal components associated with the reliably detected devices.
The three modules are executed alternatingly until convergence.
Since the channel matrix becomes sparser as the SIC iterations proceed, the pilot overhead can be far smaller than that of
the conventional spatial-domain joint ADD and CE (JADCE) scheme based on (\ref{Eq1}) only \cite{Ke_TSP'20}.

\vspace{-1mm}
\subsection{Case Study}
\label{Sec.IV-C}

\begin{figure*}[!t]
\centering
\subfloat[]{\includegraphics[width=2.9in]{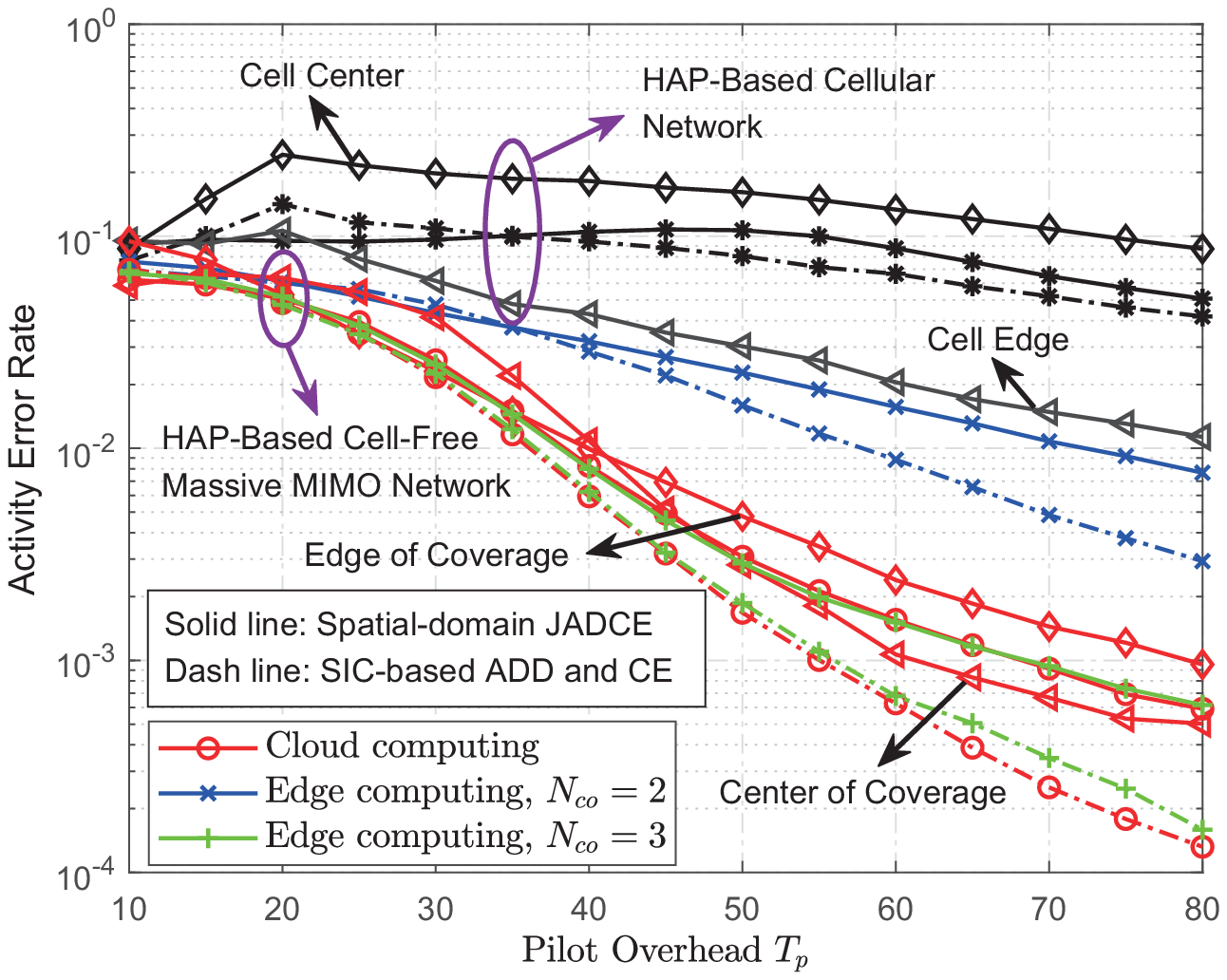}}
\hfil
\subfloat[]{\includegraphics[width=2.9in]{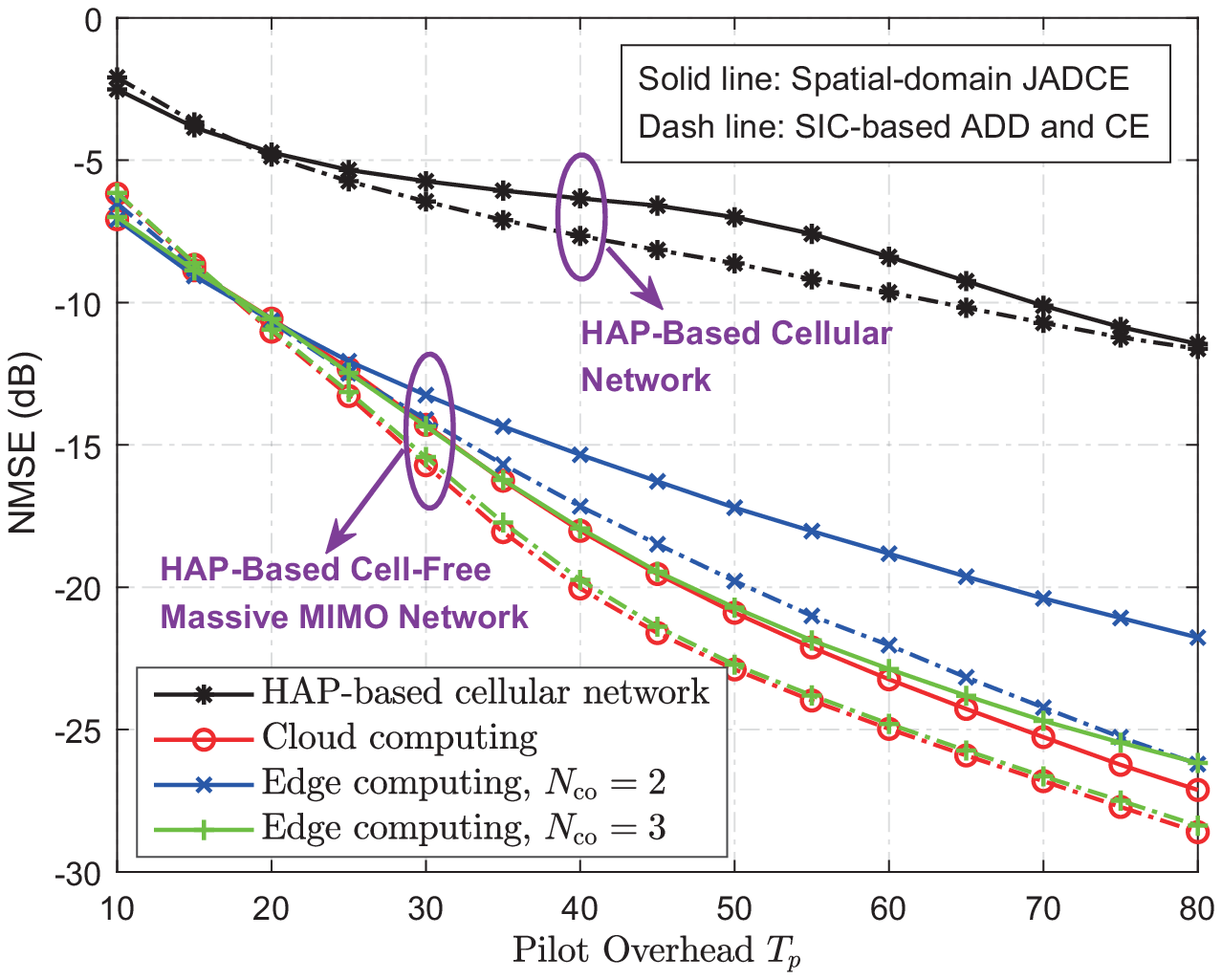}}
\caption{\small{Massive access performance comparison of the proposed solution and benchmarks.
Particularly, the ADD performances of devices distributed in the center and the edge of
the HAP coverage are indicated.}
Consider $K = 2,800$ devices are uniformly distributed in the network and $B = 7$ HAPs cooperate
in the stratosphere to serve these devices.
The number of active devices is $K_a = 140$ and the number of antennas in each HAP is $N_r = 16$.
(a) Activity error rate of ADD;
(b) Normalized mean square error (NMSE) of CE.}
\label{Fig5}
\vspace*{-3mm}
\end{figure*}

To verify the superiorities of the proposed HAP network-enabled edge computing paradigm, we compare the performance of
the traditional HAP-based cellular network, spatial-domain JADCE scheme, and cloud computing paradigm as the benchmarks.
As shown in Fig. \ref{Fig5}, the HAP-based cell-free massive MIMO network can achieve a much better
ADD and CE performance than that of the traditional HAP-based cellular network.
Meanwhile, the uniformly good service for all devices can be achieved.
Moreover, compared with the conventional spatial-domain JADCE algorithm, the adopted SIC-based ADD and CE algorithm can
dramatically reduce the access latency by further leveraging the angular domain sparsity of massive MIMO channels and
the idea of SIC.
Besides, by increasing the number of HAPs for cooperation, i.e., $N_{co}$, the performance of edge computing approaches
that of cloud computing.
Particularly, we observe that only $N_{co} = 3$ cooperative HAPs are required for edge computing to obtain
almost the same performance of cloud computing, i.e., the inter-cell interference is avoided.
This is because the channel gains from a specific active device to the far away HAPs approximate zero,
and the signals received at remote HAPs can not further improve the massive access performance.

\section{Challenges and Open Issues}
\label{Sec.V}

Despite the potentials, the research on HAP network-enabled edge computing paradigm for supporting massive IoT connectivity
is still in its infancy, where many key research issues are still open.
In this section, we discuss some main research topics and the potential solutions.

\textbf{Seamless integration with other networks:}
Satellites, HAPs, UAVs, and terrestrial BSs have various advantages and disadvantages in terms of cost, reliability,
persistence, etc.
To reap the benefits of all these infrastructures, the design of space-air-ground integrated networks will be
a promising research topic.
Here, the efficient inner-network and inter-network cooperation should be investigated for improved QoS.
Meanwhile, the major issues, such as access scheduling, resource allocation, and inter-network interference management,
become more challenging due to the multi-dimensional heterogeneity in resources, services, and network infrastructures.
In this context, software-defined networking and network function virtualization are two of the many attractive techniques
to facilitate the integrated network design~\cite{Qiu_WC'19}.

\textbf{Limited fronthaul, backhaul, and inter-network links:}
With the explosive increase of the number of devices and the emerging applications like virtual reality,
the resulting massive volume of data will pose a great challenge to the wireless fronthaul, backhaul,
and inter-network links with limited capacity.
Due to the availability of LoS propagation, the promising millimeter-wave frequency bands
and free-space optical communication can be employed for improved capacity.
This also indicates that NTN-based edge computing paradigms have the potential of achieving a lower
computation offloading latency compared to the terrestrial network-based solutions.
On the other hand, low-resolution quantization for reducing the number of bits in the conveyed signals
is another solution to tackle this challenge, but the design of the receive algorithms should take
the quantization accuracy of processing signals into account~\cite{Ke_JSAC'20}.

\textbf{Blind signal detection for grant-free massive access:}
The existing grant-free massive access schemes usually adopt training-based signal detection methods,
where CSI is first acquired based on the received pilot signals and then used for coherent data detection.
To further reduce the pilot transmission time (also one of the major components of access latency),
it is interesting to investigate the design of a blind signal detection scheme, where the channels and data
can be simultaneously estimated without utilizing any pilots.
By jointly leveraging the sporadic traffic of devices and the angular domain sparsity of massive MIMO channels,
the bilinear generalized AMP algorithm is a promising candidate to realize blind signal detection.
However, the intrinsic permutation and phase ambiguities of the estimated signals would be the new challenges~\cite{Yuan_Tcom'18}.

\textbf{Channel modeling for aerial communication links:}
The channel models adopted in current literature fail to take the unique stratospheric transmission characteristics
into account.
To understand the full potential of HAP networks, a universally agreed-upon and practically substantiated
channel model should be further studied to facilitate the performance evaluation.

\textbf{Dynamic deployment of HAP networks:}
Compared with stable terrestrial networks, the deployment of HAP networks is highly dynamic, which requires
frequent network re-configuration.
Hence, it is urgent to develop efficient self-organizing control schemes to optimize the deployment and management of
the HAP networks, like trajectory optimization, resource management, etc.
In this regard, the research on intelligent control methods, which are based on machine learning, will be super valuable.

Other open issues include but are not limited to ADD and CE in the case of time-varying channels and device activity,
wireless power transfer for prolonging the lifetime of networks, security issues, etc.
Due to the page limit, we omit their detailed discussions.
However, these research topics also deserve to be further studied in the future.

\section{Conclusions}
\label{Sec.VI}

NTN-based edge computing paradigms are promising to provide ubiquitous and low-latency computing services for
computationally-intensive IoT applications.
In massive IoT connectivity scenarios, we have investigated the limitations of existing NTN-based edge computing solutions,
in terms of the fundamental network architecture, random access procedure, and multiple access techniques.
We further proposed an attractive HAP-enabled aerial cell-free massive MIMO network to realize the NTN-based edge computing architecture,
where a grant-free massive access scheme is adopted to guarantee the reliable and low-latency connectivity to edge servers.
Compared with HAP-based cellular networks and traditional random access schemes, the HAP-enabled cell-free massive MIMO network
employing grant-free massive access scheme can effectively achieve the HAP cooperation and significantly reduce the access latency.
In addition, key challenges and open issues were discussed to provide enlightening guidance for future research directions.

%

\ifCLASSOPTIONcaptionsoff
  \newpage
\fi

\end{document}